# Mucin-inspired, high molecular weight virus binding inhibitors show biphasic binding behavior to influenza A viruses

*Matthias Wallert[1], Chuanxiong Nie[2], Parambath Anilkumar[3], Srinivas Abbina[3], Sumati Bhatia[2], Jayachandran N. Kizhakkedathu[3,4,5], Rainer Haag[2,*], Stephan Block[1,*]*

[1]Institute of Chemistry and Biochemistry, Emmy-Noether Group "Bionanointerfaces", Freie Universität Berlin, Takustr. 3, 14195 Berlin, Germany

[2]Institute of Chemistry and Biochemistry, Macromolecular Chemistry, Freie Universität Berlin, Takustr. 3, 14195 Berlin, Germany

[3]Centre for Blood Research, Life Sciences Institute, Department of Pathology and Laboratory Medicine, University of British Columbia, 2350 Health Sciences Mall, Vancouver, BC V6T 1Z3, Canada

[4]Department of Chemistry, The University of British Columbia, Vancouver, BC., Canada

[5]School of Biomedical Engineering, The University of British Columbia, Vancouver, BC., Canada

Keywords: hyperbranched polyglycerol, influenza A virus, virus binding inhibition, single particle tracking, TIRF microscopy

*Corresponding authors:

Stephan Block, e-mail: stephan.block@fu-berlin.de

Rainer Haag, email: haag@zedat.fu-berlin.de

ABSTRACT

Multivalent virus binding inhibitors are a promising new class of antivirals, preventing virus infection of cells by inhibiting the first step in the viral infection cycle –binding of viruses to the cell surface. The design of multivalent virus binding inhibitors is complex as many properties, such as inhibitor size and functionalization with virus attachment factors, have a strong impact on the inhibition efficiency. In this study, we synthesized virus binding inhibitors, the design of which has been inspired by mucins, which are naturally occurring glycosylated proteins with molecular weights in the MDa range and which show high affinity in the interaction with various viruses. Hyperbranched polyglycerols (hPG), serving as polymeric scaffolds, were functionalized with sialic acids and sulfate groups at degrees of functionalization as suggested from the structure of mucins. The molecular weights of the hPG-based inhibitors ranged between 10 and 2600 kDa, thereby hitting the size of mucins (MDa scale) and allowing for comparing the inhibition efficiency of the largest, mucin-sized inhibitor (2600 kDa) with related inhibitors of lower molecular weight. Inhibition efficiencies were determined by various methods based on the inhibition of influenza A virus (IAV) binding to lipid membranes, including an assay based on total internal reflection fluorescence (TIRF) microscopy that allows for probing the interaction of IAV with its native attachment factor, sialic acid. Potent inhibition is observed in all assays already at pM concentrations for the mucin-sized inhibitor, while decreasing the inhibitor's molecular weight also decreased its inhibition efficiency. In addition, a biphasic binding behavior of the inhibitors to IAV is observed, which is attributed to differences in the binding affinity to two IAV envelope proteins, neuraminidase and hemagglutinin.

Viral infections and resulting pandemics are a human danger and repeatedly get into the focus due to the circulation of newly emerging viruses, such as Zika, influenza, and corona viruses.[1-3] The development of antivirals to fight viral infections is often a time-consuming process and the generated drugs are often effective only for a certain period of time as mutation of the targeted virus may cause the developed antiviral to become ineffective, for example due to occurrence of resistance.[4] Hence, antivirals targeting highly conserved structures of viruses, such as envelope proteins that are involved in virus binding to the membrane of the host cell, are entering more and more the focus of current research.[5, 6] As virus binding proteins typically show a very weak affinity to their native attachment factors on the cell membrane,[7] viruses bind to cells by forming many protein-receptor interactions in parallel, thereby generating a multivalent interaction to the cell membrane.[8] This first step in the infection cycle of cells can be inhibited by addition of multivalent binding inhibitors.[5, 9] Yet the design of such compounds is complicated by the fact that many properties, such as inhibitor size and functionalization with virus attachment factors, have a strong impact on the inhibition efficiency,[10] making the process of finding inhibitor designs with high inhibition efficiency a time consuming matter.

Nevertheless, the concept of using multivalent interactions to hinder binding of infectious agents (viruses and bacteria) to cells is implemented in the biological hydrogel mucus, which covers almost all epithelia cells.[11] It forms the first defense barrier against viruses and bacteria by capturing such infectious agents before they are able to reach to the cell surface.[12] Mucus is formed by dynamic cross-linking of mucins, which are highly glycosylated proteins having molecular weights ranging between 0.1 and few MDa and bind with high specificity and selectivity to viruses while allowing other species (*e.g.*, nutrients) to pass the hydrogel.[11, 13] As mucins are effective binders of various virus species, our study aims to develop a virus binding inhibitor that is inspired by the design of mucins, allowing for potent and potentially broad band inhibition of virus binding.

In particular, we aim to generate a polymer-based multivalent virus binding inhibitor with a molecular weight hitting the million Dalton (MDa) scale, which is a value being in the middle of the weight range reported for mucins.[13] Besides mimicking the large size of mucins, such large values are also motivated by the observation that the size of a binding inhibitor has a strong impact on its efficiency to bind to viruses,[10] which led to the suggestion that the optimal size of a virus binding inhibitor is ~ one third of the size of the virus to be inhibited. As many viruses are within the size range of 30 – 200 nm, the size of the virus inhibitor should therefore be on the order of a few tens of nanometers, suggesting to use polymer scaffolds having a molecular weight hitting the MDa scale. Besides size, the functionalization of the inhibitor is also known to be an important determinant for the virus-inhibitor interaction strength. Mucins present a high amount of terminal sialic acid and sulfate moieties,[12] which are known to play a role in the binding of various viruses, motivating to functionalize the virus binding inhibitor with sialic acid and sulfate groups (at degrees of functionalization comparable to mucins) to ensure efficient binding to viruses and offer the perspective to act as a broad band virus binding inhibitor. The performance of the synthesized, mucin-inspired virus binding (MuVib) inhibitor is investigated by probing the binding inhibition of influenza A virus (IAV) strain X31 (H3N2) using various assay and by comparing its inhibition efficiency to the one of hPG inhibitors having the same functionalization but lower molecular weight.

## RESULTS AND DISCUSSION

In order to generate a high molecular weight, polymer-based, multivalent virus binding inhibitor, hyperbranched polyglycerols (hPG) are promising scaffolds in the design of such inhibitors, as hPGs have been shown to be highly biocompatible and as the synthesis of hPGs with molecular weights of several 100 kDa has recently been demonstrated, making an successful extension into the MDa range highly likely.[14, 15] The strategy for the synthesis of a MDa-sized hyperbranched polyglycerol (hPG) scaffold for use as mucin-inspired virus binding (MuVib) inhibitor is shown in Figure 1a. In addition to this MuVib inhibitor, hPG scaffolds with molecular weights ranging between 10 kDa and 500 kDa have been synthesized, which allows for investigating the effect of the inhibitor size on the inhibitor efficiency as suggested in a recent work by Vonnemann et al.[10] While hPGs up to 600 kDa can be synthesized in a single step,[16] the synthesis of the 2600 kDa hPG required an additional step, in which 600 kDa hPGs served as macroinitiator for a further polymerization (Figure 1a) to generate a mucin-sized polymer.[15] This approach yielded 4 different hPG scaffolds with molecular weights of 10 kDa, 100 kDa, 500 kDa and 2600 kDa (Table 1), exhibiting hydrodynamic diameter of approximately 6 nm (10 kDa), 9 nm (100 kDa), 13 nm (500 kDa) and 29 nm (2600 kDa), respectively.

Besides size, the functionalization of a polymer scaffold surface with chemical groups is also an important parameter for its efficiency to serve as virus inhibitor, as it determines the amount and strength of interactions formed between a virus and the inhibitor.[7] Many terminal glycans in mucins terminate with a sialic acid (SA) or sulfate group.[13] Furthermore, both chemical groups are known attachment factors for various viruses:[17] SAs are, for example, involved in the attachment of influenza virus and many corona virus strains,[18, 19] while viruses such as the vesicular stromatitis virus or herpes simplex virus bind to sulfated glycosaminoglycans.[20, 21] As these two chemical groups already enable interaction with various virus species, we simplified the complex structure of mucins by regarding only these two terminal functional groups and thus by functionalizing the hPG scaffolds with SAs and sulfate residues as shown in Figure 1b. Various values for the content of SA and sulfate groups in mucins have been reported in the literature, typically ranging between 4 -20 wt.% for sialic acid and 1.2 – 11.9 wt.% for sulfates.[22, 23] In this study, 5 mol% of the hydroxyl groups were substituted with SA and sulfate moieties (Figure 1b), which resulted in 15 wt.% SA and 5 wt.% sulfate (Table 1), being close to the middle of the ranges reported for mucins. In addition, two control polymers were prepared, carrying only one of the two moieties. All functionalized inhibitors show a negative zeta potential ranging from -22 to -27 mV (Table 1), which is caused by the incorporation the negatively charged sulfate and SA moieties. Both, unfunctionalized and functionalized hPG polymers are highly water soluble.

In a next step, we assessed the applicability of the synthesized inhibitors to hinder virus binding to cell membranes. These investigations are done by probing the inhibition of influenza A viruses (IAVs; strain X31), being a highly important representative of the viruses that are known to bind to SA or sulfate moieties. Two membrane proteins of IAV are involved in its binding to cell membranes, hemagglutinin (HA) and neuraminidase (NA). Both proteins are able to interact with SAs, but while HA is known to promote IAV attachment to the cell membrane, NA is known to possess SA cleavage activity and is thus believed to promote IAV egress (after the virus replication cycle has been completed).[24-26] According to Harris et al.[27] there are approximately 300 HA compared to 50 NA in the membrane of the influenza A virus strain X31. Recently, more evidence is accumulating that NA also contributes to the binding of IAV to the membrane.[28, 29] The interplay between HA and NA is important to pass the whole virus life cycle and gets more and more in the focus of many studies.[30] To prevent NA-mediated cleavage of SA at our synthesized binding inhibitor, a thioether bond was used to link the sialic acid to the polymer scaffold. This thioether bond cannot be enzymatically cleaved by NA. Naturally occurring SAs in mucins are O-glycosylated and can be cleaved by NA.

Initially, the inhibitor performance was investigated using the well-established hemagglutination inhibiton assay (Figure S12),[29] which takes advantage of the feature that red blood cells become crosslinked upon interaction with IAVs (called hemagglutination).[29] By the addition of virus binding inhibitors, the interaction between the virus and the red blood cells can be inhibited and the lowest inhibitor concentration at which no agglutination occurs anymore is defined as inhibition constant $k_i$. The $k_i$ of the 10 and 100 kDa inhibitors are in the low μM range, whereas the 500 kDa inhibitor is in the nM range (Table 2). The most potent inhibition was exhibited by the 2600 kDa MuVib inhibitor (31 pM; 0.1 μg/mL), irrespective if this hPG was functionalized with SA and $SO_4$ or SA alone. Sulphated hPGs showed no inhibition.

In order to further probe how the IAV-SA interaction is affected by the addition of the hPG-based inhibitors, we employed a recently introduced virus-membrane binding assay based on total internal reflection fluorescence (TIRF) microscopy (Figure 2).[31] In this assay, a supported lipid bilayer (SLB), which is supplemented with the attachment factor of the virus of interest, is formed at a glass interface and the (transient) binding of fluorescently labeled viruses to the SLB is monitored using TIRF microscopy (Figure 2a). By choosing TIRF illumination, the excitation light hits the SLB-glass interface at the critical angle of total internal reflection, which generates an evanescent wave with a penetration depth of ~ 100 – 150 nm,[32] so that fluorescence is only excited in very close vicinity of the SLB. Hence, to resolve fluorescently labeled viruses for an appreciable time period in TIRF microscopy, they have to be bound to the SLB, while unbound viruses are not visible.

For probing the IAV-SA interaction, a SLB based of POPC (96.2 wt.%, 1-palmitoyl-2-oleoyl-glycero-3-phosphocholine ) and DSPE-PEG2k (3.8 wt.%, 1 mol%, 1,2-distearoyl-sn-glycero-3-phosphoethanolamine-N-[carboxy(polyethylene glycol)-2000] (sodium salt)) was supplemented with the ganglioside GD1a (2.5 wt.%, 1 mol%), which provides the native IAV attachment factor sialic acid for the interaction studies. IAVs have been labeled with the dye rhodamine octadecyl and TIRF microscopy was used to monitor IAV attachment to and IAV release from the SA-containing SLBs (Figure 2a). Single particle tracking (SPT) applied to the resulting TIRF microscopy videos allowed for tracking the motion of single viruses interacting with the SLB with high spatial accuracy (< 15 nm) and subsequent application of the equilibrium fluctuation analysis (EFA)[33, 34] yielded information on the rate of IAV attachment to the SLB, the IAV diffusion coefficient, and residence time distribution as described previously.[31]

As the inhibitors are designed to prevent IAV binding to membranes, it is straightforward to assess their inhibition efficiency by quantifying the change of the rate of IAV attachment to the SLB as function of the inhibitor concentration (Figure 2b). According to the EFA procedure,[33, 34] this rate is extracted by calculating, for each recorded TIRF movie separately, the total number of IAVs that have been newly bound to the SLB since the beginning of the TIRF movie. Under equilibrium conditions (*i.e.*, if SLB-bound and solution-dissolved viruses are in thermodynamic equilibrium) this number, the so-called cumulative number of newly arising viruses, increases linearly with measurement time, and the slope of this function corresponds to the IAV attachment rate for the area probed.[33, 34] Addition of the inhibitors changed the slope of the cumulative number of newly arising IAVs, indicating a change of the rate of IAV attachment to the SLB that is caused by addition of inhibitors (Figure 2b). As the efficiency in inhibiting virus binding can be extracted from the change in the IAV attachment rate upon inhibitor addition,[21] in the following the relative on-rate will be given, which was calculated by normalizing the attachment rates by the value measured in absence of any inhibitor.

All functionalized hPG inhibitors showed a decrease in the relative on-rate for a sufficiently large inhibitor concentration (Figure 3). Fitting a Langmuir-type inhibition curve to the relative on-rate in this concentration range allows for determining the inhibitor concentration, at which the IAV binding rate to the SLB has been reduced by 50 %, the so-called $IC_{50}$ value (Table 2). While the 10 to 500 kDa hPG inhibitors showed $IC_{50}$ values in the μM molar range, the 2600 kDa hPG inhibitors reached pM values. This impressive value is not only due to high molecular weight of these large inhibitors, but by an improved inhibition potential, as indicated when comparing the $IC_{50}$ values expressed in mass concentrations (Table 2): Compared to the smaller inhibitors a virus inhibition is achieved for the 2600 kDa hPGs at a three order of magnitudes lower mass concentration, which is indicative for a multivalency-based enhancement of the inhibition. In addition, both SA-functionalized 2600 kDa hPGs showed comparable inhibition efficiency, *i.e.*, hPGs with SA and sulfate functionalization is not notably more effective than the inhibitor with just SA, while the one lacking SA but carrying sulfates did not show any IAV inhibition, as expected for this negative control. Unexpectedly, the non-functionalized MDa hPG also showed IAV binding inhibition with an $IC_{50}$ value of ~1000 pM. We attribute this behavior to the slightly positive zeta potential (Table S1 in SI) of this polymer, which promotes its binding to negatively charged sialic acid groups present in the SLB and thus blocks the interaction between IAVs and the SLB. This view is confirmed by the lack of any inhibition potential of $hPG_{2600}$-$SO_4$, which possesses, as all functionalized hPGs, a negative surface charge and is therefore not able to bind to the SLB. A direct interaction of the unfunctionalized polymer ($hPG_{2600}$) with the virus is unlikely, which is supported by the lack of any activity of $hPG_{2600}$ in the HAI assay. A reason for its positive zeta potential could be the incorporation of salt ions although the polymer was excessively dialyzed against water. A treatment of the polymer solution with cation exchanger before the measurement of the zeta potential did not change the slightly positive zeta potential result. All functionalized hPGs show an expected zeta potential, which verifies the measurement principle.

Furthermore, the results on the IAV binding inhibition obtained using the HAI- or TIRF-based assays are qualitatively in good agreement (Table 2), although the TIRF-based assay reports in general slightly smaller $IC_{50}$ values than the HAI-assay. Nevertheless, both assay show that the $IC_{50}$ values generally decrease with increasing inhibitor size, *i.e.*, that the inhibitors become more potent for increasing size of it scaffold. This trend is in fact expected based on the theoretical considerations of Vonnemann et al.,[10] who show that the optimal size of a virus binding inhibitor is approximately one third of the size of the virus to be inhibited. As IAV X31 typically shows spatial extensions ranging between 80 and 120 nm, an optimum in binding inhibition is expected for inhibitor diameters on the order of 30 nm, which is realized by the MuVib inhibitor ($hPG_{2600}$-SA-$SO_4$, Table 1). In this context, the decrease of the IC50 with increasing inhibitor size qualitatively matches to the predictions of Vonnemann et al.[10] A test, if the MuVib inhibitor indeed achieves optimum inhibition efficiencies was, however, not possible, as this would have required to synthesize notably larger hPG scaffolds, which was not achievable in this study.

Nevertheless, while the HAI assay provides a rather binary readout (*i.e.*, if the inhibitor concentration is sufficient to inhibit agglutination of red blood cells),[29] the changes in the IAV attachment rate measure by TIRF showed in fact a continuous and biphasic progression: Starting at small inhibitor concentrations (<< $IC_{50}$), the attachment rates first increased with increasing inhibitor concentration, followed by a saturation and a strong decrease as large inhibitor concentrations (~ $IC_{50}$). This surprising behavior indicates that the SA-functionalized hPG inhibitors actually promote IAV attachment at relatively low inhibitor concentrations, while IAV binding inhibition is observed for sufficiently large inhibitor concentrations. The first phase (increase in IAV attachment rate) is not resolvable in the HAI-assay, as the IAV concentration is chosen such that agglutination is observed in absence of the inhibitor. This means that an increase in the IAV attachment rate to the cell membranes is not resolved by HAI, as in absence of inhibitors the IAV attachment rate is already sufficiently large to achieve agglutination.

This biphasic behavior upon addition of inhibitors, which was observed for all inhibitors except for $hPG_{100}$-SA-$SO_4$ and which first increases the IAV binding rate to membranes at low inhibitor concentrations until a decrease is observed at large concentrations, was unexpected. Nevertheless, the increase of the rate of IAV attachment upon inhibitor addition resembles changes to the IAV binding properties observed upon application of neuraminidase inhibitors.[31] Hence, the biphasic behavior suggests that at relatively low concentrations the inhibitors bind first to the NA and therefore act as NA inhibitors, while binding to HA and thus binding inhibition requires higher inhibitor concentrations. This interpretation is consistent with measurements on the affinity of the envelope proteins to sialosides,[35, 36] showing that the dissociation constant $K_D$ of the NA-SA interaction is 3 order of magnitude smaller than the one of the HA-SA interaction (~ µM versus ~ mM, respectively) and supporting the hypothesis that also the inhibitors exhibit higher affinity (lower $K_D$ value) towards NA than HA.

If the hypothesis is correct that the inhibitors bind preferentially to NA at relatively low inhibitor concentrations and to HA at sufficiently large ones, one would expect to see inhibitor concentration-dependent changes of the IAV off-rate and valency distribution, as both IAV binding properties strongly depend on the functional balance between HA and NA.[31] In order to test this hypothesis, we extracted the off-rate and valency distributions as recently described for monovalent inhibitors.[31] In brief, here we make use of the fact that the diffusion coefficient of SLB-bound IAVs, *D*, decreases with increasing valency (= number of bound GD1a gangliosides), so that the changes of the IAV valency distribution also induce change the distribution of IAV diffusion coefficients. Although the exact relationship connecting IAV valency and diffusion coefficient has not yet been resolved, all theoretical models imply that both properties are connected by a monotonously decreasing function, *i.e.*, large values of the diffusion coefficient correspond to a small value of the average valency and *vice versa*.[37] Furthermore, the measurement of the IAV diffusion coefficient also allows to deconvolute the IAV residence time distribution from valency effects, yielding valency-resolved off-rate distributions.[31]

In absence of any inhibitor (open circles, Figure 4a), the observed off-rate distribution is dominated by a decrease for increasing apparent average valency $1/D$, but also shows a peak structure, leading to elevated off-rates at intermediate average valencies ($1/D \sim 8$ s/µm²). This finding matches to previous observations,[31] in which it is shown that this peak structure is caused by the opposing functionalities of HA and NA and vanishes upon application of NA inhibitors such as zanamivir. Furthermore, the addition of zanamivir also increased the rate of IAV attachment and both effects together yielded a strong increase in the number of bound IAVs. Interestingly, exactly the same behavior is observed here at relatively low inhibitor concentrations, at which the relative IAV on-rate increases (Figure 3) and the IAV off-rate decreases with increasing inhibitor concentration (open triangles and squares, Figure 4a). Hence, at relatively low concentrations, the inhibitors behave like a NA inhibitor with respect to the induced changes in IAV attachment and off-rate distribution. Additional evidence is provided by the diffusion coefficient distributions (Figure 4b), which showed a shift from large to small values (indicating an increase in average binding valency; "NA-like inhibition") upon addition of small amounts of the inhibitor (<< $IC_{50}$) and a shift from small to large diffusion coefficient values (indicating an decrease in average binding valency; "HA inhibition") upon addition of small amounts of the inhibitor (>> $IC_{50}$).

Furthermore, the phenomenon of the increased IAV attachment at small inhibitor concentrations (<< $IC_{50}$) was further investigated by binding tests to MDCK-II cells. Here, labelled IAVs were mixed with different concentration of the inhibitor $hPG_{2600}$-SA-$SO_4$ at room temperature for 45 min. Afterwards, the virus-inhibitor mixture was incubated with MDCK-II cells for 2 h on ice. Non-bound IAVs were removed by washing twice with PBS buffer, while bound IAVs remained at the cell surface. Using confocal laser scanning microscopy Z-wide images of the cell layer were taken and stacked to visualize all viruses at the cell surface, followed by quantifying the number of bound viruses using ImageJ.[38] The number of MDCK-II cell-bound IAVs (Figure 5) showed the same dependence of the inhibitor concentration as the relative on-rate determined using TIRF microscopy (Figure 3), verifying an increased IAV attachment at low concentrations and a decrease in IAV attachment at higher concentrations. The sulfated derivative ($hPG_{2600}$-$SO_4$, 10 nM) was used as negative control and shows the same number of attached viruses as the lowest concentration ($10^{-4}$ pM) of $hPG_{2600}$-SA-$SO_4$. Hence, the biphasic change in the IAV binding behavior upon inhibitor addition is observed for IAVs interacting with attachment factor-equipped SLBs and with cell membranes, which further supports our view that these inhibitors behave as NA-like inhibitors at low inhibitor concentrations and as HA inhibitors at high concentrations.

## CONCLUSIONS

In this study, new hyperbranched polyglycerol (hPG)-based virus binding inhibitors have been synthesized, the functionalization of which (sialic acids and sulfate groups) have been inspired from mucins. The molecular weights of the synthesized inhibitors covers the range between 10 and 2600 kDa, thereby hitting the size of mucins (MDa scale) and allowing for comparing the inhibition efficiency of the mucin-inspired virus binding (MuVib) inhibitor (2600 kDa molecular weight) with related inhibitors of lower molecular weight. The efficiency of the synthesized compounds to inhibit the binding of a common respiratory virus, a H3N2 influenza A virus (IAV), was assessed using a cell binding and the hemagglutination inhibition assay, both of which provide information about the minimum inhibitor concentration that needed to inhibit IAV binding to membranes. Furthermore, a recently developed TIRF-based assay was employed to quantify, how inhibitor addition modified the multivalent interaction arising between IAVs and their native attachment factor, sialic acids. Surprisingly, the IAV binding to membranes showed a biphasic behavior for increasing inhibitor concentrations: At relatively low inhibitor concentrations, the IAV binding first increased with increasing inhibitor concentrations, followed by a second regime at higher inhibitor concentrations, at which the expected decrease in IAV binding was observed for increasing inhibitor concentrations. The results of the TIRF-based assay and the cell binding assay indicate that the inhibitors binds at low inhibitor concentrations preferentially to the IAV envelope protein neuraminidase (NA), leading to an enhancement of the IAV binding to lipid membranes, while at larger inhibitor concentrations, it binds additionally to another IAV envelope protein, hemagglutinin (HA), causing IAV binding inhibition. Potent inhibition of IAV binding (HA inhibition) is observed in all these assays already at pM concentrations for the MuVib inhibitor (2600 kDa molecular weight), while decreasing the molecular weight of the inhibitors decreased their inhibition efficiency (increase in their $IC_{50}$ value), which is in qualitative agreement with the prediction by a recent study of Vonnemann et al.[10]

ACKNOWLEDGEMENTS

This work was supported by the German Research Foundation (BL1514/1-1 and project A6 within the CRC 765) and by the Focus Area NanoScale (Freie Universität Berlin). The authors thank the core facility BioSupraMol and the Macromolecular Hub (CBR) for the use of their research facilities. JNK acknowledges the funding by Canadian Institutes of Health Research (CIHR), Natural Sciences and Engineering Council of Canada (NSERC) and Canada Foundation for Innovation (CFI). JNK holds a Career Investigator Scholar award from the Michael Smith Foundation for Health Research (MSFHR). SA acknowledges a MSFHR postdoctoral fellowship.

ASSOCIATED CONTENT

**Supporting Information**

The Supporting Information is available free of charge on the ACS Publications website at DOI: 10.1021/xxx.

Details on the synthesis and characterization of all inhibitors are given in Note S1 and S2, respectively. A GPC measurement of the hPG2600 scaffold is given in Figure S1. NMR spectra of the synthesized compounds are shown in Figures S2, S3, and S6 – S11, while Figures S4 and S5 summarize IR spectra of the compounds at different stages of the functionalization process.

Figure S12 gives an example for the HAI assay. Furthermore, the properties of the hPG scaffolds and inhibitors are summarized in Table S1 and S2, respectively. (PDF)

## AUTHOR INFORMATION

### Corresponding Authors

*E-mail: stephan.block@fu-berlin.de.

*E-mail: haag@zedat.fu-berlin.de.

### Notes

The authors declare no competing financial interest.

MATERIALS AND METHODS

**Synthesis of high-molecular weight (MDa) hPG** – The synthesis of the MDa hPG was perfomed in two steps.[15] First, a macroinitiator was synthesized in a heterogenous reaction mixture in dioxane. Dry trimethylolpropane (120 mg, 0.89 mmol, 1.0 eq.) was partially deprotonated (30 % OH) with potassium methoxide (67 µL, 0.27 mmol, 0.3 eq., 25 % in methanol) in argon atmosphere at 60 °C for 30 min. After the addition of 24 mL dioxane (dry) the turbid mixture was heated to 100 °C. Glycidol (12 mL, 0.18 mol, 201 eq.) was slowly added (0.5 ml/h) via syringe pump into the reaction mixture. The polymer was purified by removing the dioxane, precipitation as methanolic solution in acetone and dialysis against water in regenerated cellulose membrane (10 kDa MWCO). The resulted hPG (Mn: 600 kDa, *Đ* 1.1) was obtained with a yield of 93.6 %. This hPG was used as macroinitiator to grow the polymer further. 2.5 g (0.034 mol, total OH groups) of the lyophilized polymer was dissolved in dry DMF (35 mL). The polymer was partially deprotonated with the addition of potassium hydride in oil (30 wt.%) (80 µg, 272 µL, 2.0 µmol). The temperature was increased to 100 °C and glycidol (25 mL, 0.37 mol) were added with a rate of 0.9 mL/h. After precipitation in acetone and dialysis against water in regenerated cellulose membrane (50 kDa MWCO) the resulted molecular weight was 2.6 MDa with a *Đ* of 1.4.

**Mesylation and azidation** – Both reactions were performed sequentialy in one pot. 1300 mg hPG (2.6 MDa) (0.88 mol OH to be functionalized) was dissolved in dry DMF (18 mL). The mesylation was done with methanesulfonyl chloride (201 mg, 1.8mmol, 2.0 eq.) in the present of triethyl amine (306 µL, 2.2 mmol, 2.5 mol eq.) at room temperature for 16 h. Afterwards the azidation was directly performed by the addition of sodium azide (456 mg, 7.0 mmol, 8 eq.) at 80 °C for 20 h under argon atmosphere. Purification was done by dialysis against water (2 kDa

MWCO, benzoylated). The product was analyzed by $^{1}$H NMR and elementary analysis to determine the amount of introduced azide groups (see Figure S2 in SI).

**Sulfation** – The azidated hPG (400 mg, 0.27 mmol OH to be functionalized, 1.0 eq.) was sulfated using sulfur trioxide pyridine complex (55.9 mg, 0.4 mmol, 1.3 eq.) in dry DMF (12 mL) at 60 °C for 20 h under argon atmosphere. After the reaction time was over the pH value was increased to pH 10 by the addition of sodium hydroxide solution (0.3 mol/L). The product was dialyzed against sodium hydroxide solution (0.3 mol/L), 10 wt. % NaCl and water. The yield of the sulfation was analyzed by elementary analysis. Further, a $^{1}$H NMR of the product was performed (see Figure S3 in SI).

**Click protected propagylated sialic acid by CuAAC** – The sulfated polymer $hPG_{2600}$-$N_3$-$SO_4$ (150 mg, 0.11 mmol OH to be functionalized, 1.0 eq.) was mixed with acetyl protected propagylated sialic acid (72.9 mg, 1.3 mmol, 1.2 eq.) in DMF. Copper sulfate pentahydrate (11.1 mg, 0.04mmol, 0.4 eq.) and sodium-L- ascorbate (88.3 mg, 0.4 mmol, 4 eq.) were dissolved separately in a small amount of water and combined afterwards. The solution was transferred to the polymer solution. The reaction was performed at 50 °C for 20 h under argon atmosphere. Complete conversion was verified by the disappearance of the azide band (2100 cm$^{-1}$) with infrared spectroscopy (see Figure S4 in SI). The pH value was increased to pH 10 with sodium hydroxide solution (2 mol/L) to cleave the acetyl protection of the sialic acid within 2 h. EDTA disodium salt (14.0 mg, 0.04 mmol, 0.4 eq.) was added to improve the removal of copper ions during dialysis against water. The purified polymer was analyzed by $^{1}$H NMR (see Figure S6 in SI) and elementary analysis.

The synthesized virus binding inhibitors were characterized in regard to size with dynamic light scattering (DLS) and in respect of surface charge by zeta potential measurements.

**DLS** - The hydrodynamic diameter was measured by dynamic light scattering at a concentration of 1 mg/mL in PBS buffer using Zetasizer Nano series ($\lambda$ = 532 nm) from Malvern Panalytical (Kassel, Germany). Disposable cuvettes (ZEN0040) from Brand (Wertheim, Germany) out of polystyrene were used. Before the measurement all samples were filtered through a 0.2 µm Minisart RC 15 syringe filter from Satorius (Göttingen, Germany). Temperature equilibration was done for 1 min at 25 °C. The measurements were performed for 10 scans each 15 s in back scattering mode (173 °). The stated values result from at least 3 measurements.

**Zeta-potential** - The surface charge was investigated by zeta-potential measurement with Zetasizer Nano series ($\lambda$ = 532 nm) using folded capillary zeta cells (DTS 1070) from Malvern Panalytical (Kassel, Germany). The sample concentration was 1 mg/mL in 10 mM phosphate buffer solution (0.411 g/L $NA_2HPO_4$, 0.178 g/mL $KH_2PO_4$, pH 7.4). All samples were filtered through a 0.2 µm Minisart RC 15 syringe filter from Satorius (Göttingen, Germany). 5 measurements with 10 scans (each 15 s) were done to obtain the zeta potential based on the Smoluchowski model.

The virus binding performance was investigated by two independent methods: hemagglutination inhibition (HAI) assay and total internal reflection fluorescence (TIRF) microscopy.

**Hemagglutination inhibition (HAI) assay** – The inhibitors were two-fold diluted with PBS in a v-shaped microtiter plate. Afterwards, 2 HA X31 virus were transferred to each inhibitor dilution. After 30 min incubation time at room temperature 50 µL of 1 % red blood cell solution (Robert Koch-Institute, Berlin) was added to each well. Then, it was incubated for 60 min at room temperature before the read out was done. The lowest concentration where the sedimentation of red blood cells was still inhibited by hemagglutination represents the inhibitor constant $k_i$.

**Total internal reflection fluorescent (TIRF) microscopy** – Single virus tracking was done on a supported lipid bilayer (SLB), which represents the cell surface. IAV bind in a multivalent fashion to incorporated GD1a gangliosides (1 mol %). Extruded vesicles out of POPC and 1 mol% GD1a were exposed in an aqueous solution (0.33 mg/mL) on a cleaned glass, absorb, deform and form a cohesive supported lipid bilayer by rupturing on the glass surface. Sialic acids on the GD1a are the natural target of IAVs. After 10 min excess vesicles were removed by washing with PBS buffer. Then, IAV solution with or without containing binding inhibitor was injected. Through excitation in TIRF mode, an evanescent light beam (white light + m-cherry filter) penetrates approximately 100 nm in the solution and reaches therefore just bound viruses (~ 100 nm) on the membrane. This ensures that rhodamine (R18) labeled viruses in the evanescent region get excited and emit light, non-bound labeled virus in the solution are not visible. Videos with 0.11 fps were taken to visualize the virus attachment, diffusion and detachment on the GD1a receptor containing SLB. The videos are analyzed by homemade MATLAB scripts using equilibrium fluctuation analysis (EFA).

**IAV binding to MDCK-II cells** – The cultured MDCK II cells are seeded in 8-well confocal slides and cultured for 1-2 days till confluency. 100 µL X31 solution (protein content: 0.36mg/mL, $1.1 \times 10^{11}$ particles/mL) was incubated with 2µL of 20µM DiO (in ethanol, D4929 in Merck, Darmstadt, Germany) for 30 min in dark. The free dyes were removed by spin column then. A mixture of 90 µL inhibitor (in PBS) and 10 µL labelled virus was added after 45 min at room temperature on the cell layer (culture medium was removed before by washing with PBS twice). After an incubation for 2 h on ice, non-binding viruses were removed by washing twice with PBS. The cell nucleus was labelled with Hoechst 23358 and the cells fixed with 4% paraformaldehyde. Finally, Z-wide images were taken with confocal laser scanning microscopy from the whole cell layer (30 images, step size 0.4 μm). The number of binding viruses was determined from the stacked images using ImageJ.

REFERENCES

1. Hoffmann, M.; Kleine-Weber, H.; Schroeder, S.; Kruger, N.; Herrler, T.; Erichsen, S.; Schiergens, T. S.; Herrler, G.; Wu, N. H.; Nitsche, A.; Muller, M. A.; Drosten, C.; Pohlmann, S., SARS-CoV-2 Cell Entry Depends on ACE2 and TMPRSS2 and Is Blocked by a Clinically Proven Protease Inhibitor. *Cell* **2020**.
2. Ventura, C. V.; Maia, M.; Bravo-Filho, V.; Góis, A. L.; Belfort, R., Zika virus in Brazil and macular atrophy in a child with microcephaly. *The Lancet* **2016,** *387* (10015), 228.
3. Neumann, G.; Noda, T.; Kawaoka, Y., Emergence and pandemic potential of swine-origin H1N1 influenza virus. *Nature* **2009,** *459* (7249), 931-939.
4. De Clercq, E.; Li, G., Approved Antiviral Drugs over the Past 50 Years. *Clin Microbiol Rev* **2016,** *29* (3), 695-747.
5. Fasting, C.; Schalley, C. A.; Weber, M.; Seitz, O.; Hecht, S.; Koksch, B.; Dernedde, J.; Graf, C.; Knapp, E. W.; Haag, R., Multivalency as a chemical organization and action principle. *Angew Chem Int Ed Engl* **2012,** *51* (42), 10472-98.
6. Kiessling, L. L.; Gestwicki, J. E.; Strong, L. E., Synthetic multivalent ligands in the exploration of cell-surface interactions. *Current opinion in chemical biology* **2000,** *4* (6), 696-703.
7. Mammen, M.; Choi, S. K.; Whitesides, G. M., Polyvalent Interactions in Biological Systems: Implications for Design and Use of Multivalent Ligands and Inhibitors. *Angew Chem Int Ed Engl* **1998,** *37* (20), 2754-2794.
8. Mulder, A.; Huskens, J.; Reinhoudt, D. N., Multivalency in supramolecular chemistry and nanofabrication. *Org Biomol Chem* **2004,** *2* (23), 3409-24.
9. Marsh, M.; Helenius, A., Virus entry: open sesame. *Cell* **2006,** *124* (4), 729-40.
10. Vonnemann, J.; Liese, S.; Kuehne, C.; Ludwig, K.; Dernedde, J.; Bottcher, C.; Netz, R. R.; Haag, R., Size dependence of steric shielding and multivalency effects for globular binding inhibitors. *J Am Chem Soc* **2015,** *137* (7), 2572-9.
11. Witten, J.; Samad, T.; Ribbeck, K., Selective permeability of mucus barriers. *Current opinion in biotechnology* **2018,** *52*, 124-133.
12. Fahy, J. V.; Dickey, B. F., Airway mucus function and dysfunction. *N Engl J Med* **2010,** *363* (23), 2233-47.
13. Bansil, R.; Turner, B. S., Mucin structure, aggregation, physiological functions and biomedical applications. *Current Opinion in Colloid & Interface Science* **2006,** *11* (2-3), 164-170.
14. Calderón, M.; Quadir, M. A.; Sharma, S. K.; Haag, R., Dendritic polyglycerols for biomedical applications. *Advanced materials* **2010,** *22* (2), 190-218.
15. Anilkumar, P.; Lawson, T. B.; Abbina, S.; Mäkelä, J. T.; Sabatelle, R. C.; Takeuchi, L. E.; Snyder, B. D.; Grinstaff, M. W.; Kizhakkedathu, J. N., Mega macromolecules as single molecule lubricants for hard and soft surfaces. *Nature Communications* **2020,** *11* (1), 1-9.
16. Imram ul-Hag, M.; Shenoi, R. A.; Brooks, D. E.; Kizhakkedathu, J. N., Solvent-assisted anionic ring opening polymerization of glycidol: Toward medium and high molecular weight hyperbranched polyglycerols. *Journal of Polymer Science Part a-Polymer Chemistry* **2013,** *51* (12), 2614-2621.

17. Cone, R. A., Barrier properties of mucus. *Adv Drug Deliv Rev* **2009,** *61* (2), 75-85.
18. Sauter, N. K.; Hanson, J. E.; Glick, G. D.; Brown, J. H.; Crowther, R. L.; Park, S. J.; Skehel, J. J.; Wiley, D. C., Binding of influenza virus hemagglutinin to analogs of its cell-surface receptor, sialic acid: analysis by proton nuclear magnetic resonance spectroscopy and X-ray crystallography. *Biochemistry* **1992,** *31* (40), 9609-21.
19. Tortorici, M. A.; Walls, A. C.; Lang, Y.; Wang, C.; Li, Z.; Koerhuis, D.; Boons, G. J.; Bosch, B. J.; Rey, F. A.; de Groot, R. J.; Veesler, D., Structural basis for human coronavirus attachment to sialic acid receptors. *Nat Struct Mol Biol* **2019,** *26* (6), 481-489.
20. Guibinga, G. H.; Miyanohara, A.; Esko, J. D.; Friedmann, T., Cell surface heparan sulfate is a receptor for attachment of envelope protein-free retrovirus-like particles and VSV-G pseudotyped MLV-derived retrovirus vectors to target cells. *Molecular Therapy* **2002,** *5* (5), 538-546.
21. Peerboom, N.; Schmidt, E.; Trybala, E.; Block, S.; Bergstrom, T.; Pace, H. P.; Bally, M., Cell Membrane Derived Platform To Study Virus Binding Kinetics and Diffusion with Single Particle Sensitivity. *ACS Infect Dis* **2018,** *4* (6), 944-953.
22. Loomis, R. E.; Prakobphol, A.; Levine, M. J.; Reddy, M. S.; Jones, P. C., Biochemical and biophysical comparison of two mucins from human submandibular-sublingual saliva. *Arch Biochem Biophys* **1987,** *258* (2), 452-64.
23. Bolscher, J.; Veerman, E.; Van Nieuw Amerongen, A.; Tulp, A.; Verwoerd, D., Distinct populations of high-M(r) mucins secreted by different human salivary glands discriminated by density-gradient electrophoresis. *Biochem J* **1995,** *309 ( Pt 3)*, 801-6.
24. de Vries, E.; Du, W.; Guo, H.; de Haan, C. A. M., Influenza A Virus Hemagglutinin-Neuraminidase-Receptor Balance: Preserving Virus Motility. *Trends Microbiol* **2020,** *28* (1), 57-67.
25. Hamming, P. E.; Overeem, N. J.; Huskens, J., Influenza as a molecular walker. *Chemical science* **2019,** *11* (1), 27-36.
26. Wagner, R.; Matrosovich, M.; Klenk, H. D., Functional balance between haemagglutinin and neuraminidase in influenza virus infections. *Rev Med Virol* **2002,** *12* (3), 159-66.
27. Harris, A.; Cardone, G.; Winkler, D. C.; Heymann, J. B.; Brecher, M.; White, J. M.; Steven, A. C., Influenza virus pleiomorphy characterized by cryoelectron tomography. *Proc Natl Acad Sci U S A* **2006,** *103* (50), 19123-7.
28. Yang, J.; Liu, S.; Du, L.; Jiang, S., A new role of neuraminidase (NA) in the influenza virus life cycle: implication for developing NA inhibitors with novel mechanism of action. *Rev Med Virol* **2016,** *26* (4), 242-50.
29. McAuley, J. L.; Gilbertson, B. P.; Trifkovic, S.; Brown, L. E.; McKimm-Breschkin, J. L., Influenza Virus Neuraminidase Structure and Functions. *Front Microbiol* **2019,** *10*, 39.
30. Guo, H.; Rabouw, H.; Slomp, A.; Dai, M.; van der Vegt, F.; van Lent, J. W. M.; McBride, R.; Paulson, J. C.; de Groot, R. J.; van Kuppeveld, F. J. M.; de Vries, E.; de Haan, C. A. M., Kinetic analysis of the influenza A virus HA/NA balance reveals contribution of NA to virus-receptor binding and NA-dependent rolling on receptor-containing surfaces. *PLoS Pathog* **2018,** *14* (8), e1007233.

31. Müller, M.; Lauster, D.; Wildenauer, H. H. K.; Herrmann, A.; Block, S., Mobility-Based Quantification of Multivalent Virus-Receptor Interactions: New Insights Into Influenza A Virus Binding Mode. *Nano Lett* **2019,** *19* (3), 1875-1882.
32. Axelrod, D., Total internal reflection fluorescence microscopy. *Methods Cell Biol* **1989,** *30*, 245-70.
33. Lee, D. W.; Hsu, H. L.; Bacon, K. B.; Daniel, S., Image Restoration and Analysis of Influenza Virions Binding to Membrane Receptors Reveal Adhesion-Strengthening Kinetics. *PLoS One* **2016,** *11* (10), e0163437.
34. Bally, M.; Gunnarsson, A.; Svensson, L.; Larson, G.; Zhdanov, V. P.; Hook, F., Interaction of single viruslike particles with vesicles containing glycosphingolipids. *Phys Rev Lett* **2011,** *107* (18), 188103.
35. Zhu, X.; McBride, R.; Nycholat, C. M.; Yu, W.; Paulson, J. C.; Wilson, I. A., Influenza virus neuraminidases with reduced enzymatic activity that avidly bind sialic Acid receptors. *J Virol* **2012,** *86* (24), 13371-83.
36. Amaro, R. E.; Ieong, P. U.; Huber, G.; Dommer, A.; Steven, A. C.; Bush, R. M.; Durrant, J. D.; Votapka, L. W., A Computational Assay that Explores the Hemagglutinin/Neuraminidase Functional Balance Reveals the Neuraminidase Secondary Site as a Novel Anti-Influenza Target. *ACS Cent Sci* **2018,** *4* (11), 1570-1577.
37. Block, S., Brownian Motion at Lipid Membranes: A Comparison of Hydrodynamic Models Describing and Experiments Quantifying Diffusion within Lipid Bilayers. *Biomolecules* **2018,** *8* (2).
38. Schneider, C. A.; Rasband, W. S.; Eliceiri, K. W., NIH Image to ImageJ: 25 years of image analysis. *Nat Methods* **2012,** *9* (7), 671-5.

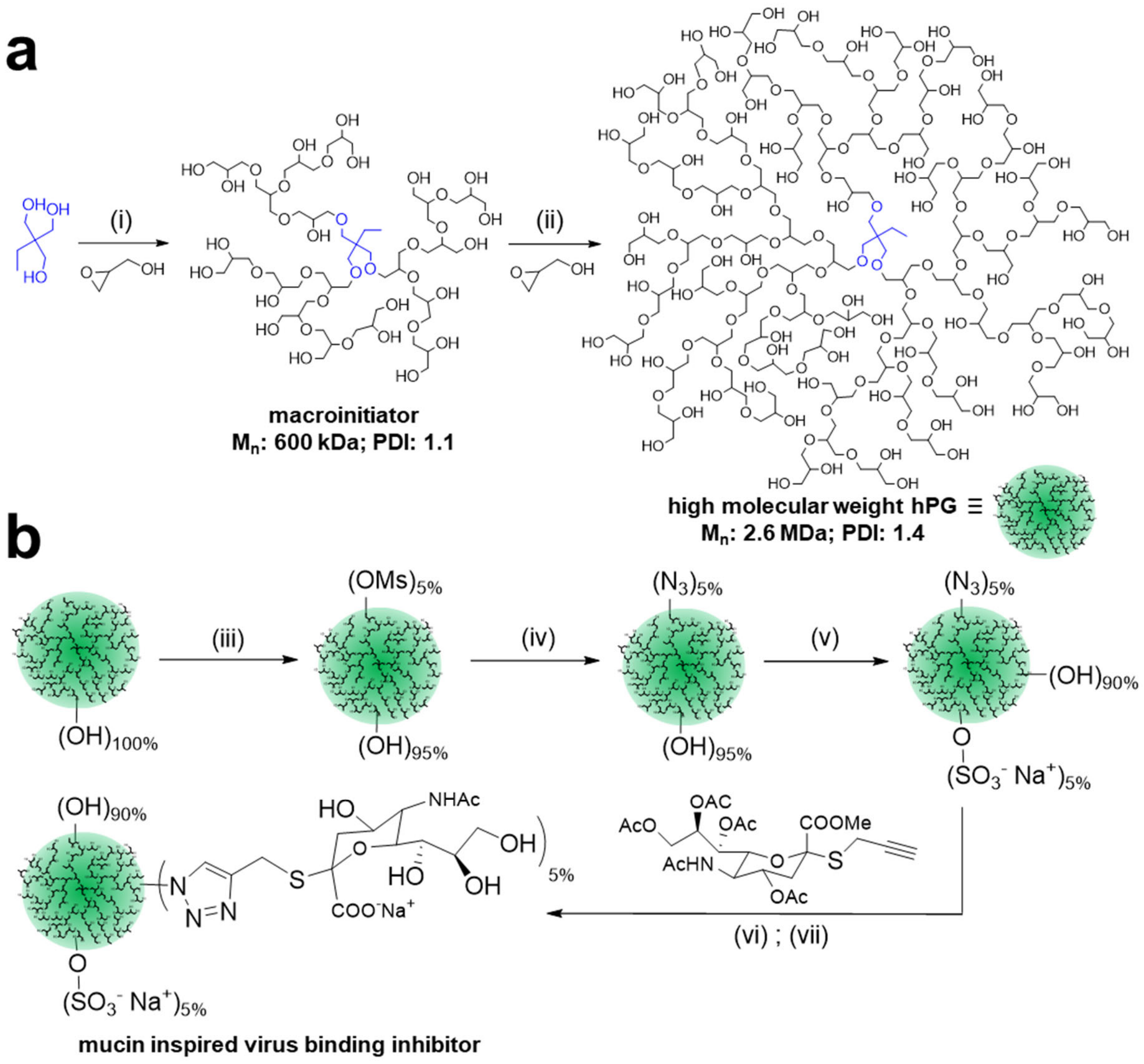

**Figure 1**. Syntheses of mucin-inspired virus binding (MuVib) inhibitor based on high molecular weight hPG. Polymerization of the MDa hPG in two steps (a), potassium methoxide in dry 1,4 dioxane at 95 °C and glycidol addition (0.5 mL/h) (i), potassium hydride in dry DMF at 95 °C and glycidol addition (0.9 mL/h) (ii). Functionalization of the MDa hPG, all reactions were carried out in dry DMF under argon atmosphere (b), mesylation with mesyl chloride and triethylamine at room temperature for 20 h (iii), azidation using sodium azide at 80 °C for 20 h (iv), sulfation with sulfur trioxide pyridine complex at 60 °C for 20 h (v), copper-catalyzed alkyne-azide cycloaddition using copper sulfate pentahydrate, sodium ascorbate at 50 °C for 20 h (vi), deprotection with sodium hydroxide at room temperature for 2h (vii).

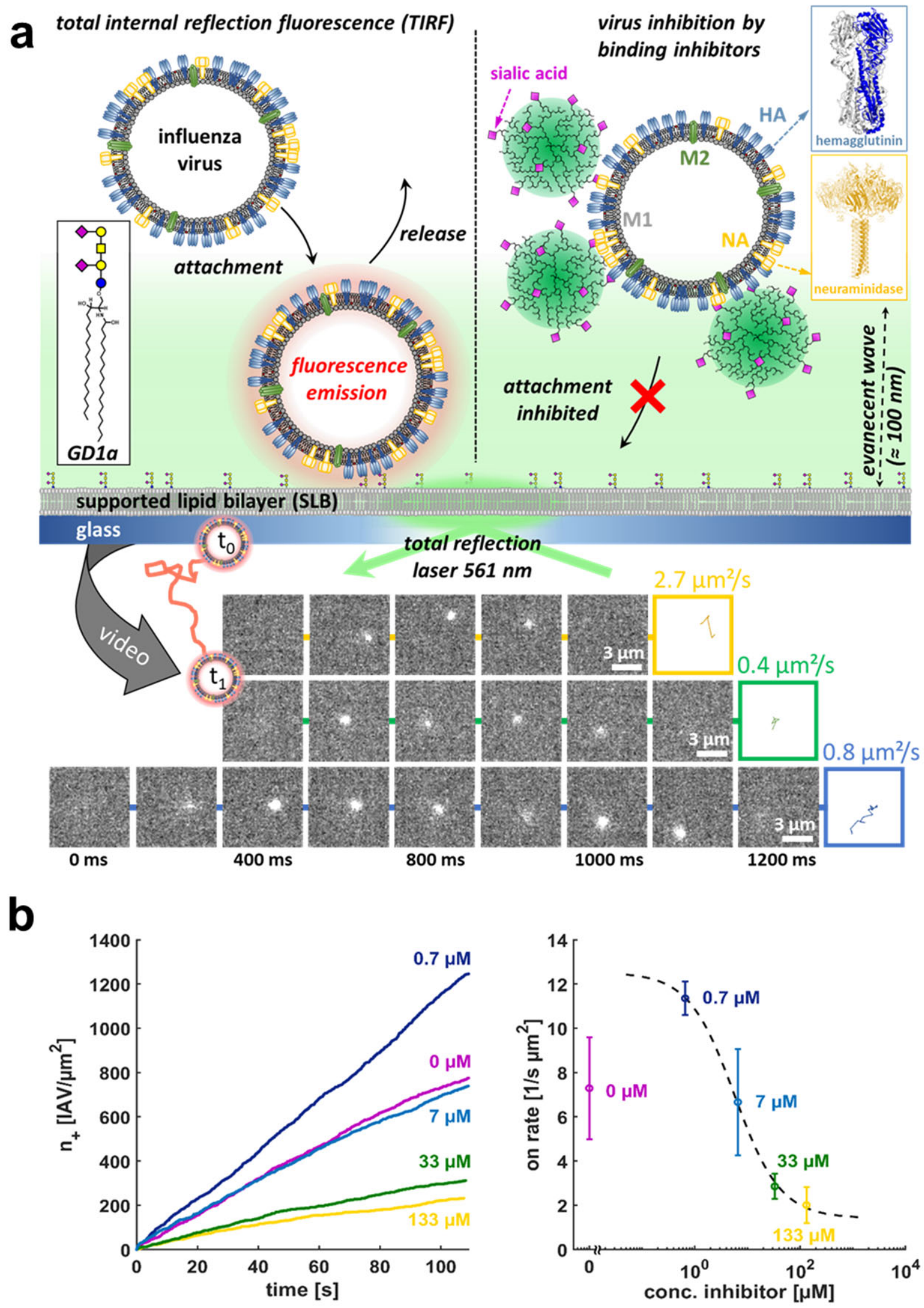

**Figure 2**. Experimental setup used to study IAV X31 binding to sialic acids (SAs) and the modification of this interaction upon addition of hPG inhibitors. (**a**) Supported lipid bilayers (SLBs) containing the sialic acid-presenting ganglioside GD1a were used as artificial cell membranes. Fluorescently labeled IAVs (R18 dye incorporated in the virus envelope) bind to sialic acids exposed by the SLB, a process which is followed by TIRF imaging (always performed at room temperature; 20 °C). (**b**) The cumulative number of newly arriving IAVs, as calculated using the EFA procedure, increases linearly with time. The slope of these traces is proportional to the rate of IAV attachment to the SA-containing SLB and allows for determining IAV binding inhibition upon addition of hPG inhibitors. The plot shows a representative example using the inhibitor $hPG_{10}$-SA-$SO_4$.

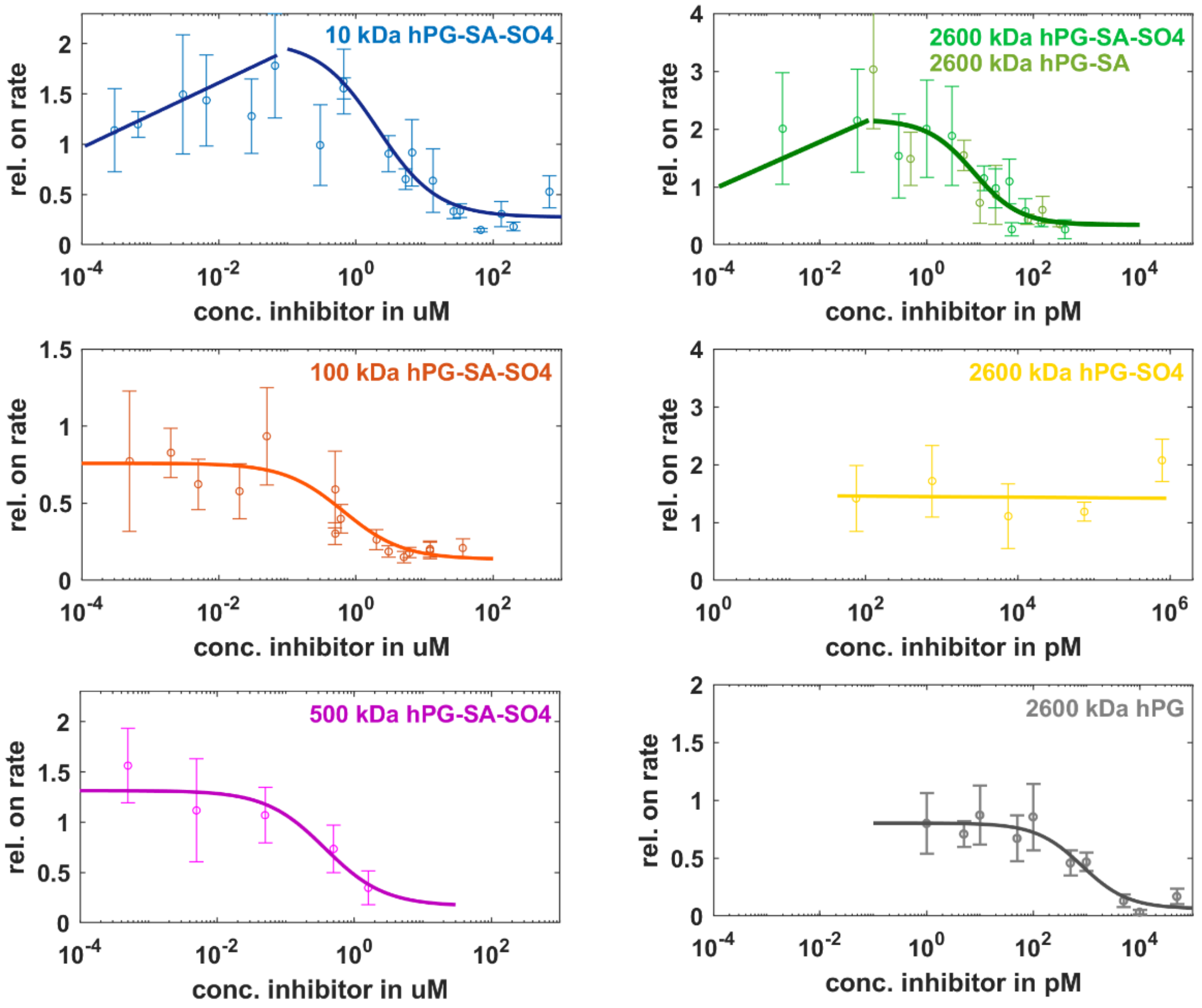

**Figure 3**. Change in IAV attachment rates upon hPG inhibitor addition. Shown is the relative on-rate, which was calculated by normalizing the IAV attachment rate at a given hPG inhibitor concentration by the IAV attachment rate in absence of the inhibitor. Linear solid lines are to guide the eye; sigmoidal solid lines are fits of a Langmuir-type inhibition model to the data. Symbols indicate average values ± standard deviation of at least 5 measurements.

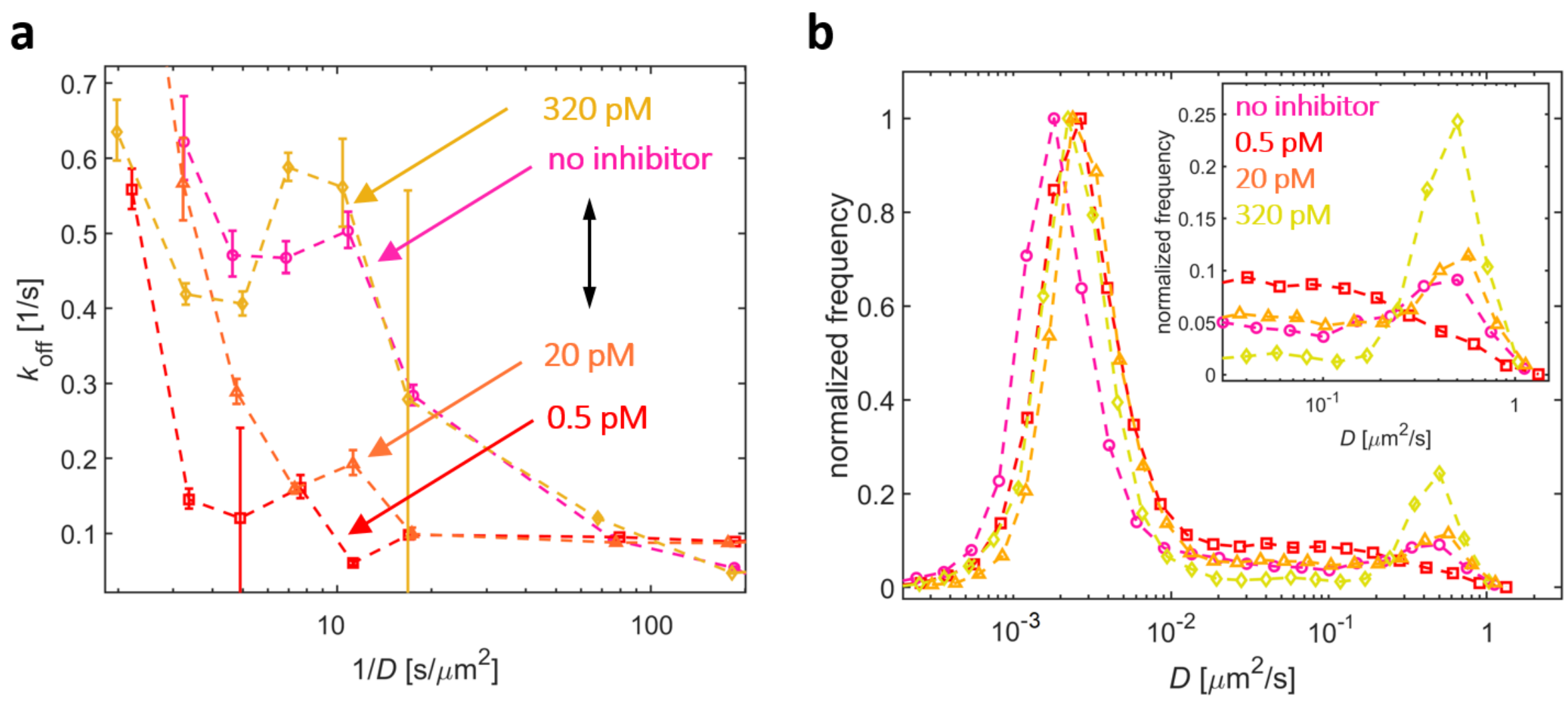

**Figure 4**. Changes of the IAV off-rate (**a**) and diffusion coefficient distribution (**b**) upon addition of the MuVib (hPG$_{2600}$-SA-SO$_4$) inhibitor. The biphasic binding behavior is also observed in these distribution, as addition of small amounts of the inhibitor (≤ 20 pM) leads to a decrease in the off-rate distribution and a shift of the diffusion coefficient distribution from large to small values (indicating an increase in average binding valency), while addition of larger amounts of the inhibitor (> 20 pM) restores the off-rate distribution and shifts the diffusion coefficient distribution from small to large values (indicating an decrease in average binding valency).

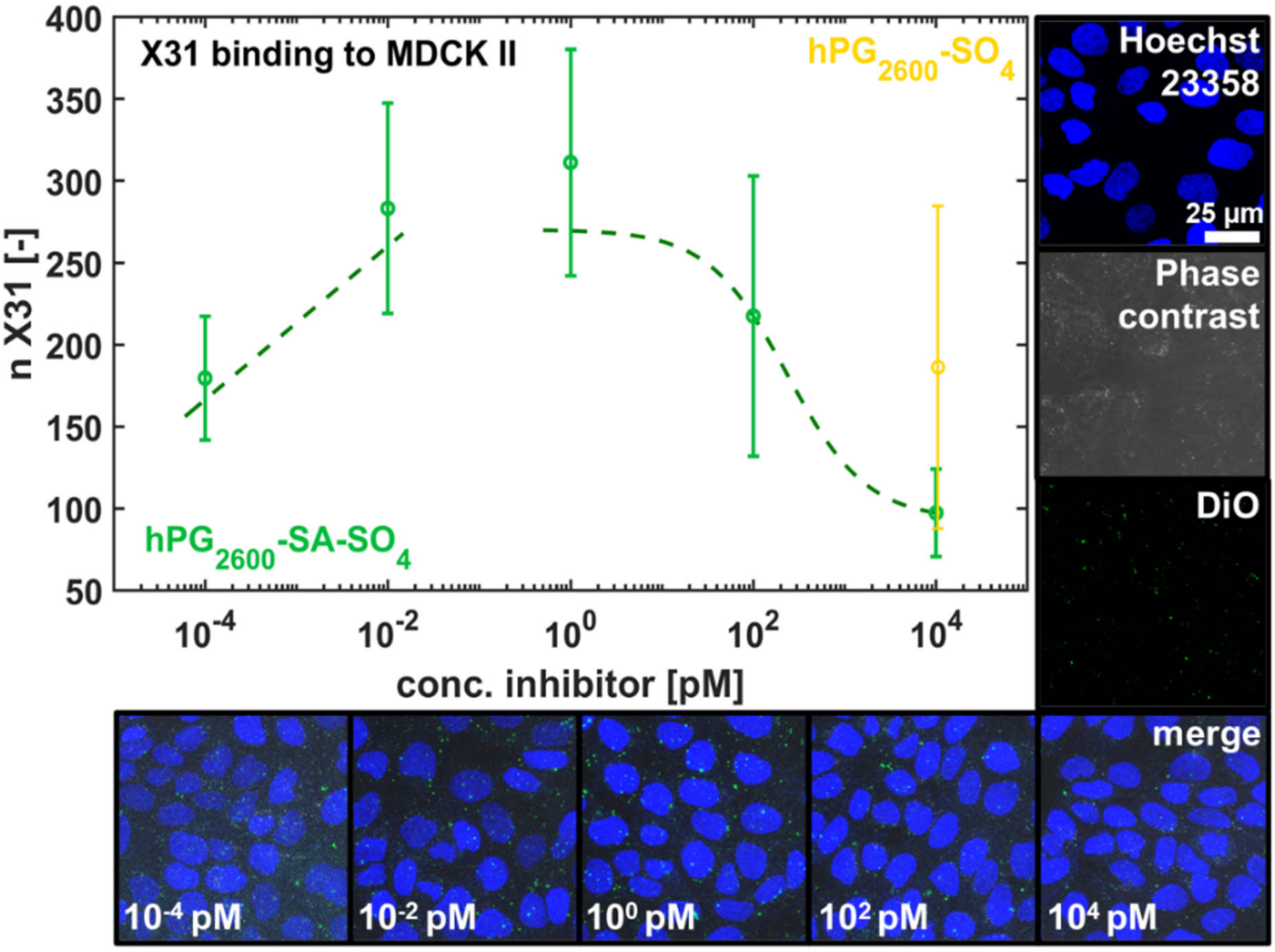

**Figure 5**. Inhibition of IAV X31 binding to MDCK II cells upon addition of the MuVib inhibitor ($hPG_{2600}$-SA-$SO_4$, green symbols) and of the negative control $hPG_{2600}$-$SO_4$ applied at a high concentration (yellow symbol). The cellular assay shows the same biphasic change of IAV binding behavior as observed in the TIRF-based binding assay. IAV were labeled with DiO (green structures in the cell images), while the MDCK II cells were labeled using Hoechst 23358 (blue structures). Symbols indicate average values ± standard deviation of 4 measurements.

**Table 1**. Properties of synthesized virus binding inhibitors based on hPG.

| | d[a] | ζ-pot.[b] | $SO_4$[c] | | SA[c] | | $n_{sulfate}$[d] | $n_{SA}$[d] |
|---|---|---|---|---|---|---|---|---|
| | [nm] | [mV] | [mol%] | [wt.%] | [mol%] | [wt.%] | [-] | [-] |
| $hPG_{10}$-SA-$SO_4$ | 4.9 ± 4.2 | -26.3 ± 0.9 | 7.2 | 6.2 | 7.0 | 19.3 | 7 | 7 |
| $hPG_{100}$-SA-$SO_4$ | 12.0 ± 7.2 | -22.4 ± 1.0 | 5.0 | 4.7 | 5.5 | 16.5 | 68 | 75 |
| $hPG_{500}$-SA-$SO_4$ | 16.3 ± 9.7 | -24.3 ± 0.8 | 3.9 | 3.7 | 5.7 | 17.1 | 268 | 392 |
| $hPG_{2600}$-SA-$SO_4$ | 28.1 ± 17.1 | -26.8 ± 1.1 | 5.0 | 4.7 | 5.6 | 16.7 | 1724 | 1931 |
| $hPG_{2600}$-SA | 29.2 ± 15.3 | -23.4 ± 0.5 | - | - | 5.6 | 17.7 | - | 1931 |
| $hPG_{2600}$-$SO_4$ | 20.8 ± 11.4 | -17.1 ± 0.9 | 4.3 | 5.3 | - | - | 1482 | - |
| $hPG_{2600}$ | 28.9 ± 14.9 | 9.5 ± 0.5 | - | - | - | - | - | - |

a) Hydrodynamic diameter obtained by DLS 1 mg/mL in PBS; [b)] Zeta-potential measurement in 10 mM phosphate buffer (pH 7.4); [c)] amount of converted hydroxyl groups in percent, determined by elementary analysis, degree of sialic acid based on fully conversion of azide moieties; [d)] calculated number of sialic acid and sulfate per polymer based on $M_n$.

**Table 2**. Inhibitor performance in comparison of HAI and TIRF measurements

| | HAI $k_i$ | | TIRF $IC_{50}$ | |
|---|---|---|---|---|
| | [mol/L] | [µg/mL] | [mol/L] | [µg/mL] |
| $hPG_{10}$-SA-$SO_4$ | $82 \times 10^{-6}$ | 1214 | $2 \times 10^{-6}$ | 30 |
| $hPG_{100}$-SA-$SO_4$ | $5 \times 10^{-6}$ | 685 | $0.6 \times 10^{-6}$ | 82 |
| $hPG_{500}$-SA-$SO_4$ | $50 \times 10^{-9}$ | 35 | $0.4 \times 10^{-6}$ | 279 |
| $hPG_{2600}$-SA-$SO_4$ | $31 \times 10^{-12}$ | 0.1 | $30 \times 10^{-12}$ | 0.1 |
| $hPG_{2600}$-SA | $31 \times 10^{-12}$ | 0.1 | $3 \times 10^{-12}$ | 0.01 |
| $hPG_{2600}$-$SO_4$ | no inh. | no inh. | no inh. | no inh. |